\documentclass[sigconf]{acmart}

\usepackage[ruled,vlined]{algorithm2e}
\usepackage{tabularx}
\usepackage{url}
\usepackage{caption}
\usepackage{graphicx}
\usepackage{booktabs}
\usepackage{amsmath}
\usepackage{enumitem}
\usepackage{subcaption}
\usepackage{comment}
\usepackage{xcolor}
\usepackage{filecontents}
\usepackage{romannum}
\usepackage{seqsplit}
\usepackage{balance}

\newcommand{\hash}[1]{{\ttfamily\seqsplit{#1}}}
\definecolor{darkblue}{RGB}{25, 25, 112}

\def\sysnameabbr{\texttt{DiscoverPath}}

\AtBeginDocument{%
  \providecommand\BibTeX{{%
    \normalfont B\kern-0.5em{\scshape i\kern-0.25em b}\kern-0.8em\TeX}}}

\title{DiscoverPath: A Knowledge Refinement and Retrieval System for Interdisciplinarity on Biomedical Research}

\copyrightyear{2023} 
\acmYear{2023} 
\setcopyright{acmlicensed}\acmConference[CIKM '23]{Proceedings of the 32nd ACM International Conference on Information and Knowledge Management}{October 21--25, 2023}{Birmingham, United Kingdom}
\acmBooktitle{Proceedings of the 32nd ACM International Conference on Information and Knowledge Management (CIKM '23), October 21--25, 2023, Birmingham, United Kingdom}
\acmPrice{15.00}
\acmDOI{10.1145/3583780.3614739}
\acmISBN{979-8-4007-0124-5/23/10}


\begin{document}

\settopmatter{authorsperrow=4}

\author{Yu-Neng Chuang}
\email{ynchuang@rice.edu}
\affiliation{%
  \institution{Rice University}
  \country{}
}

\author{Guanchu Wang}
\email{guanchu.wang@rice.edu}
\affiliation{%
  \institution{Rice University}
  \country{}
}

\author{Chia-Yuan Chang}
\email{cychang@tamu.edu }
\affiliation{%
  \institution{Texas A$\&$M University}
  \country{}
}

\author{Kwei-Herng Lai}
\email{khlai@rice.edu}
\affiliation{%
  \institution{Rice University}
  \country{}
}

\author{Daochen Zha}
\email{daochen.zha@rice.edu}
\affiliation{%
  \institution{Rice University}
  \country{}
}

\author{Ruixiang Tang}
\email{ruixiang.Tang@rice.edu}
\affiliation{%
  \institution{Rice University}
  \country{}
}

\author{Fan Yang}
\email{fyang@rice.edu}
\affiliation{%
  \institution{Rice University}
  \country{}
}

\author{Alfredo Costilla Reyes}
\email{acostillar@rice.edu}
\affiliation{%
  \institution{Rice University}
  \country{}
}

\author{Kaixiong Zhou}
\email{kz34@rice.edu}
\affiliation{%
  \institution{Rice University}
  \country{}
}

\author{Xiaoqian Jiang}
\email{xiaoqian.Jiang@uth.tmc.edu}
\affiliation{%
  \institution{UTHealth at Houston}
  \country{}
}

\author{Xia Hu}
\email{xia.hu@rice.edu}
\affiliation{%
  \institution{Rice University}
  \country{}
}
\renewcommand{\shortauthors}{Yu-Neng Chuang, et al.}

\begin{abstract}

The exponential growth in scholarly publications necessitates advanced tools for efficient article retrieval, especially in interdisciplinary fields where diverse terminologies are used to describe similar research. Traditional keyword-based search engines often fall short in assisting users who may not be familiar with specific terminologies. To address this, we present a knowledge graph based paper search engine for biomedical research to enhance the user experience in discovering relevant queries and articles. The system, dubbed \sysnameabbr{}, employs Named Entity Recognition (NER) and part-of-speech (POS) tagging to extract terminologies and relationships from article abstracts to create a KG. To reduce information overload, \sysnameabbr{} presents users with a focused subgraph containing the queried entity and its neighboring nodes and incorporates a query recommendation system enabling users to iteratively refine their queries. The system is equipped with an accessible Graphical User Interface that provides an intuitive visualization of the KG, query recommendations, and detailed article information, enabling efficient article retrieval, thus fostering interdisciplinary knowledge exploration. 
\sysnameabbr{} is open-sourced at \href{https://github.com/ynchuang/DiscoverPath}{\color{darkblue} \texttt{https://github.com/ynchuang/DiscoverPath}} with a demo video at \href{https://youtu.be/xcDzBl7jp-s}{\color{darkblue} \texttt{Youtube}}.
\end{abstract}



\maketitle




\vspace{-0.1cm}
\section{Introduction}
\begin{figure}[t!]
\setlength{\abovecaptionskip}{0mm}
\setlength{\belowcaptionskip}{-3mm}
\begin{center}    \includegraphics[width=0.47\textwidth]{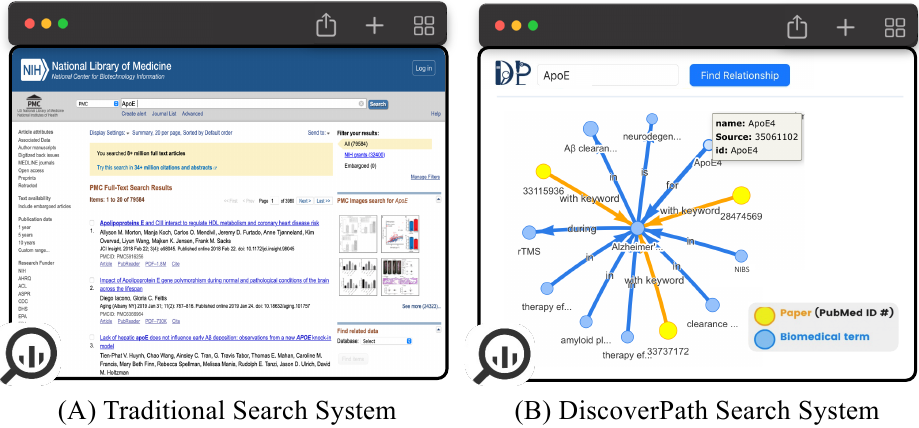}
    \vspace{0.2cm}
    \caption{Left: the traditional paper search engine PubMed. Right: we can identify the important query ``ApoE4'' in the proteomics domain with a knowledge graph since ApoE4 is connected with Alzheimer’s disease.}
    \label{fig:kg1}
    \vspace{-0.35cm}
\end{center}
\end{figure}

With an increasing number of papers being published over the years, search engines like Google Scholar and PubMed search engine\footnote{\url{https://pubmed.ncbi.nlm.nih.gov/}} have become common tools for researchers and practitioners to identify papers of interest. For example, if a genetics professional wants to explore recent advances in Alzheimer's disease research outcomes, she can use relevant phrases such as ``Alzheimer's disease'' as search queries. The search engine will then match the queries with the content of the articles and generate a ranked list of relevant articles, as illustrated in the left-hand side of Figure~\ref{fig:kg1}.

However, in many cases, users may encounter difficulty in effectively identifying suitable search queries, especially in interdisciplinary fields where researchers from different backgrounds tend to employ diverse terminologies to describe similar research. Take, for instance, a genetics expert exploring Alzheimer's disease research within the proteomics domain. If she is not familiar with proteomics-related terminologies, she may struggle to grasp important queries like Apolipoprotein E4 (ApoE4)\footnote{ApoE4 is a protein that plays a significant role in mammalian fat metabolism and is closely associated with Alzheimer's disease research within the proteomics domain.}, making it difficult to retrieve the relevant articles. This knowledge gap often leads to users spending significant amounts of time filtering out irrelevant information, thereby impeding the advancement of interdisciplinary research.

\begin{figure*}[t!]
\begin{center}
    \includegraphics[width=0.75\textwidth]{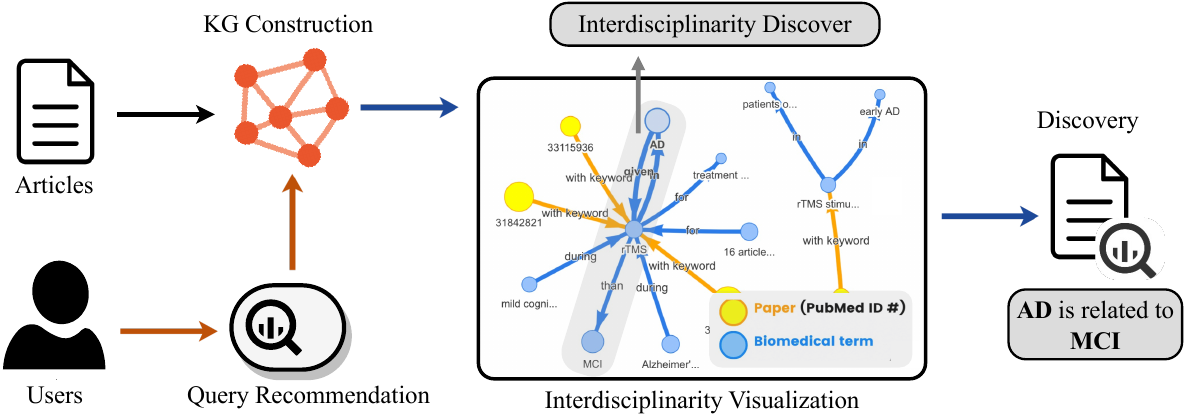}
    \caption{An overview of \sysnameabbr{} for interdisciplinarity knowledge discovery. The KG construction module extracts key terminologies from research articles and converts them into a KG. Starting from an initial query, users progressively refine and query based on suggested related queries from the query recommendation while examining subgraphs generated by the interdisciplinarity visualization. Throughout this process, users can access detailed information on each article of interest.}
    \label{fig:pip_over}
    \vspace{-0.3cm}
\end{center}
\end{figure*}

Unlike traditional search engines, a knowledge graph (KG)~\cite{hogan2021knowledge} offers a more concise and effective way to grasp the interconnectedness of key terminologies. A KG consists of nodes, which represent entities, and edges, which comprehensively represent relationships between those entities from real-world assertions~\cite{dong2023active, chuang2022mitigating}. By utilizing a KG to represent the intricate relationships between terminologies, users can easily discover novel queries or relevant information when retrieving articles of interest~\cite{dong2023hierarchy, chang2020query}.
For example, a user can readily identify the ApoE4 protein as a relevant query if it is connected to Alzheimer's disease in the KG, as shown in the screenshot of our system on the right of Figure~\ref{fig:kg1}. Several recent advancements~\cite{tan2019deep,nicholson2020constructing} have shown the effectiveness of KG-based search engines, but none of them focus on interdisciplinary exploration. Motivated by the concepts above, we ask: \emph{\textbf{Can we build a KG-based paper search engine with a user-friendly interface for exploring and discovering relevant queries and articles, especially in interdisciplinary fields?}}

However, it is non-trivial to build such a system due to several challenges. \emph{(\romannum{1}) How to construct a KG encompassing various terminologies in research articles?} Detecting the specific terminologies and their corresponding relationships can be challenging, presenting a significant obstacle in constructing a KG. \emph{(\romannum{2}) How to enable users to identify and retrieve relevant articles efficiently?} The constructed KG may include an overwhelming number of entities and relations, making it challenging for users to identify the relevant information efficiently. \emph{(\romannum{3}) How to design the Graphical User Interface (GUI)?} The system's intended users may lack a background in KG or computer science, making it challenging to design a GUI that is accessible and user-friendly for individuals in other disciplines.

To address the challenges and promote interdisciplinary knowledge exploration, we demonstrate \sysnameabbr{}, a KG-based retrieval system designed for biomedical research. This system aims to assist biomedical researchers in dynamically refining their queries and effectively retrieving articles. \sysnameabbr{} is featured for:

\begin{itemize}[leftmargin=*]
    \item \emph{A KG specifically for biomedical research.} We extract essential terminologies and relationships from the article abstracts to convert textual data into a KG. Additionally, each article is treated as an entity to aid in identifying the relevant articles.
    
    \item \emph{A query recommendation system.} To mitigate information overload, we present users with a focused subgraph that only includes the queried entity and its neighboring nodes rather than displaying the entire KG. Moreover, we develop a query recommendation system, which allows users to iteratively modify their queries and observe the resulting subgraphs, facilitating a progressive ``path'' toward discovering the most relevant articles.
    
    \item \emph{A user-friendly GUI.} We offer visualization of knowledge graphs, query recommendations, and detailed information display for each discovered article. An online demo is readily available.
\end{itemize}

\vspace{-0.25cm}
\section{DiscoverPath System}
Figure~\ref{fig:pip_over} presents an overview of \sysnameabbr{}. The system consists of three modules: (\romannum{1}) a KG construction module responsible for extracting terminologies and relations from articles and constructing the KG, (\romannum{2}) a query recommendation system module that offers relevant query suggestions to refine user queries, and (\romannum{3}) an interdisciplinarity visualization module that provides a GUI.

\subsection{Knowledge Graph Construction}
We aim to construct a KG with ``entities'' and ``relations'' to illustrate the knowledge extracted from the abstract of the articles structurally. Typically, a KG comprises several triplets extracted from the sentences in the format of head entity, relation, and tail entities. Motivated by~\cite{stewart2019icdm, wu2023medical}, we construct the KG in two steps: (1) recognizing biomedical-related terminologies as entities and (2) extracting verbs and prepositions as relations between each biomedical-related term. Specifically, we adopt name entity recognition (NER) methods from BERN~\cite{lee2020biobert} to extract the potential biomedical-related entities in the abstracts, where BERN is a neural biomedical entity recognition tool trained on articles from PubMed. Then, we extract the relation from the raw abstract utilizing the pre-trained part-of-speech tagging model from NeuralCoref~\cite{srinivasa2018natural}. 

We have presented the constructed KG to biomedical scientists to validate the correctness of the generated knowledge graphs. The feedback from the scientists indicates that the constructed KG is consistent with the relationships of the articles.

\subsection{Query Recommendation System}

The query recommendation system aims to suggest queries for refinement, where the recommended queries can be either an article or a terminology. We train a recommendation model to learn implicit relationships among the article-to-terminology networks to generate the queries effectively.

The model is trained based on a collection of articles $(\mathcal{A}, w, w')\in \mathcal{D}_\mathcal{A}$, where $w$ and $w'$ denote the relevant and irrelevant terminology of article $\mathcal{A}$, respectively.
Instantiated by~\cite{tan2023collaborative,zhou2023adaptive,tan2023bring}, the objective is to learn the relationship between an article $\mathcal{A}$ and a positive article $\mathcal{A}^+$ that shares the same terminology as $\mathcal{A}$, and a negative article $\mathcal{A}^-$ that contains the irrelevant terminology $w'$ of $\mathcal{A}$. In the article-to-terminology network, the relations of each article are generally composed of its relevant terminologies and its paper relationship.
The optimization criterion is to find an embedding matrix $\Theta$ that maximizes the joint likelihood as follows:
\begin{align}
    \notag \text{Objective} := & \ln p(\Theta|>_\mathcal{A}) ~ \propto~
    \ln \prod_{\mathcal{D}_\mathcal{A}}  p( w >_{\mathcal{A}} w' | \Theta)p(\Theta) \\
    = &\sum_{\mathcal{D}_\mathcal{A}} \ln  \sigma\Big(\big\langle \Theta_{w\mathcal{A}}, (\Theta_{w\mathcal{A}^{+}} - \Theta_{w'\mathcal{A}^{-}}) \big\rangle \Big)- \lambda \left\Vert \Theta \right\Vert^2,
    \label{eq:obj}
\end{align}
where $w >_{\mathcal{A}} w'$ denotes that article $\mathcal{A}$ related to terminology $w$ over terminology $w'$ and $\lambda$ is a regularization parameter. For every article $\mathcal{A}$, in Equation~\ref{eq:obj}, $\Theta_{w\mathcal{A}}$ represents the combination of an article $\mathcal{A}$ and relevant terminology $w$; and $\Theta_{w\mathcal{A}} = \Theta_{\mathcal{A}} + \Theta_{w}$.

\subsection{Interdisciplinarity Visualization}
\begin{figure*}[t!]
\setlength{\abovecaptionskip}{1mm}
\setlength{\belowcaptionskip}{-3mm}
\begin{center}
    \includegraphics[width=0.8\textwidth]{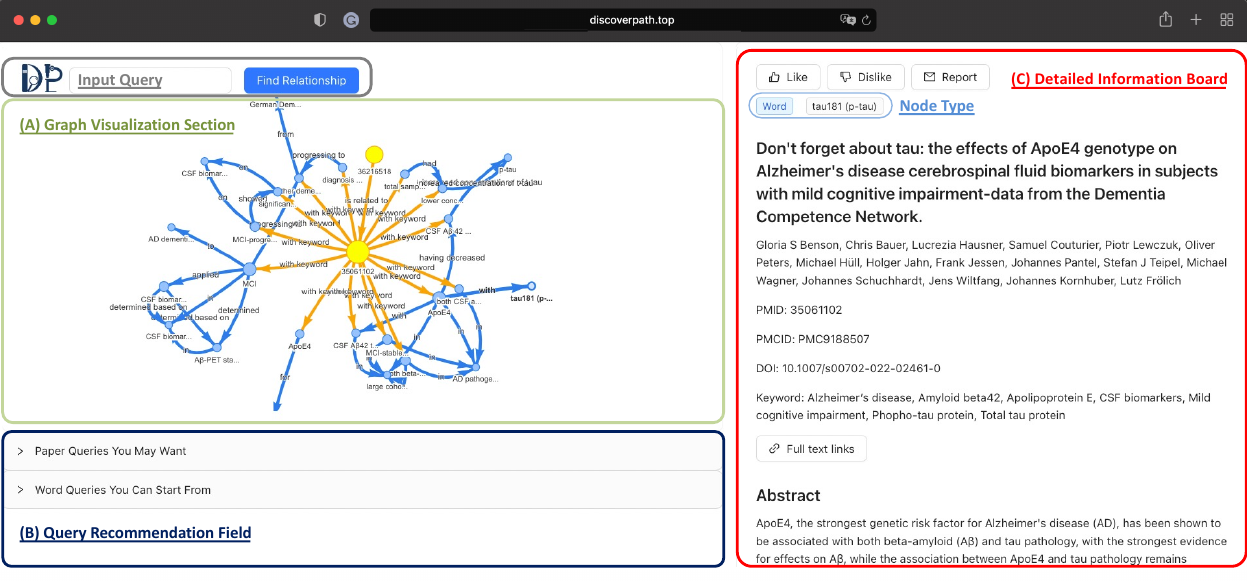}
    \caption{The GUI of \sysnameabbr{} system.}
    \label{fig:frame}
    \vspace{-0.2cm}
\end{center}
\end{figure*}

The GUI of \sysnameabbr{}, as depicted in Figure~\ref{fig:frame}, encompasses three interfaces: (\romannum{1}) a graph visualization section that presents a visualization of the KG, (\romannum{2}) a query recommendation field equipped with a search bar for suggesting queries, and (\romannum{3}) a detailed information board for displaying information of selected entities.

\subsubsection{Graph Visualization Section \& detailed information board}

The KG constructed for biomedical research is extensive and complex, making it challenging for users to easily explore entity information at first glance when presented as a static image~\cite{archambault2011difference, burch2014flip}. To tackle this obstacle, we present the KG through interactive visualization. Users can effortlessly zoom in on a particular KG element and drag entities to access further insights via pop-up windows. Additionally, double-clicking an entity unveils relevant details on the right-hand side information board. By incorporating these interactive features, \sysnameabbr{} offers a more user-friendly and intuitive way to explore interdisciplinary knowledge.


\subsubsection{Query Recommendation Field}

During interdisciplinary exploration, initial queries provided by users may not always be precise, resulting in retrieval results that do not fully meet their expectations. We use query recommendations in \sysnameabbr{} to address this issue. The query recommendations are displayed in two folders to interact with users, helping them refine their queries and obtain personalized and informative KG in the graph visualization section. The query recommendations can enhance the efficiency and effectiveness of interdisciplinary exploration.



\subsection{Interface Implementation Details}
\sysnameabbr{} is implemented based on a client-server architecture consisting of a frontend interaction, a backend platform, and a graph database. More details are shown as follows:

\begin{itemize}[leftmargin=*]
    \item \textbf{Frontend (client).} The frontend interface is built upon React~\cite{react} framework, handling the communication with the backend via HTTP API endpoints. 

    \item \textbf{Backend (API server).} The backend framework Flask~\cite{flask} offers a flexible structure for API development and database connection configuration. The backend processes retrieve data and provide query recommendation algorithms for better search results. 

    \item \textbf{Graph database.} Neo4j~\cite{neo4j} graph database is used to store and manage the data. The graph database can efficiently manage complex and highly correlated data like KG.
\end{itemize}

Note that our system is not restricted to the biomedical domain and can be easily implemented in other research areas. We can also use other LLM-based KG construction methods~\cite{zhu2022multi, ye2022generative} and recommendation algorithms~\cite{tan2020learning, liu2022ipr, chen2016query, wang2019kgat, tan2020learning} in the implementation.


\section{User Pipeline and Case Study}

In this section, we present the user's perspective workflow of \sysnameabbr{} with a case study focused on Alzheimer's disease.



\subsection{Interdisciplinarity Discovery Pipeline}
\label{sec:pipeline}

The user pipeline of \sysnameabbr{} for interdisciplinarity discovery is illustrated as follows:


\begin{itemize}[leftmargin=*]
    \item \textbf{Step 1: Input an initial query.} Users are required to provide their queries in the search bar. After clicking the search bottom, \sysnameabbr{} can retrieve the related subgraph for personalized exploration based on their initial queries.

    \item \textbf{Step 2: Observe retrieved KG.} After receiving the knowledge graphs provided by \sysnameabbr{}, users can deeply explore the relevant information for interdisciplinary discovery with a GUI and visualization. Recommended relevant papers are also listed in \sysnameabbr{} for wider exploration.
    
    \item \textbf{Step 3: Explore additional knowledge:} If users need more information, \sysnameabbr{} provides query recommendations to help the users amend their initial queries to retrieve better match-specific requirements. Users can go back to Step 1 with refined queries and move to Step 2 for advanced exploration. Step 3 is highly encouraged for whom are not familiar with the appropriate search queries in specific disciplines.
    
\end{itemize}

\subsection{Case Study of \sysnameabbr{}}

Given Alzheimer's prominence as the fifth leading cause of death in adults over 65, we employ the query ``\emph{Alzheimer}'' to showcase our system. In the visualized KG, yellow nodes denote articles' PubMed IDs, blue nodes signify query-related terms, and edges indicate node relationships. Paths link articles and terminology, while terminology-article-terminology paths uncover potential interdisciplinary connections. Readers are encouraged to search with other AD-related queries, including \emph{MCI}, \emph{amyloid}, \emph{cognitive}, and \emph{brain}. In Figure~\ref{fig:case1}, the following insights can be discerned: 


\begin{figure}[t]
\setlength{\abovecaptionskip}{1mm}
\setlength{\belowcaptionskip}{-3mm}
\begin{center}
    \includegraphics[width=0.4\textwidth]{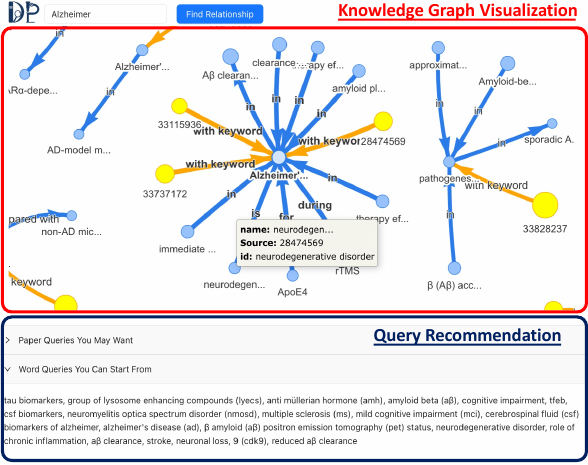}
    \vspace{0.1cm}
    \caption{Case study on Alzheimer's disease. Users input the query "Alzheimer" in the search bar to retrieve the results.}
    \label{fig:case1}
    \vspace{-0.25cm}
\end{center}
\end{figure}

\begin{itemize}[itemsep=1pt,leftmargin=*]

    \item \textbf{Terminologies Related with \emph{Alzheimer}.} 
    The neighbor terminology nodes of Alzheimer are: \emph{rTMS}\footnote{\emph{Transcranial magnetic stimulation}: A noninvasive form of brain stimulation.}, \emph{Apo-E~}\footnote{\emph{Apolipoprotein-E}: A protein involved in the metabolism of fats of mammals.}, \emph{A$\beta$ clearance\footnote{A complex process mediated by various systems and cell types.}}, etc. The edges connected to these nodes indicate the relationships of the terminologies with Alzheimer's disease. For instance, ApoE4 is the factor \emph{for} Alzheimer's disease (AD) and Alzheimer's disease \emph{is} a neurodegenerative disorder. The visualized relationships between these terminologies and Alzheimer's disease align with findings and discussions in previous works~\cite{benson2022don, marcelli2018involvement}.

    \item \textbf{Articles Related with \emph{Alzheimer}.} The neighbor article nodes of Alzheimer's disease are the papers with PMID 28474569~\cite{marcelli2018involvement}, 33737172~\cite{xu2021inhibition}, and 33115936~\cite{chu2021cognitive}. The edges linking these paper nodes reveal that Alzheimer's disease is a prevalent terminology within these papers.
     
    \item \textbf{Query Recommendation.} The recommended queries are shown beneath the KG visualization section, including ``tau biomarkers'', ``group of lysosome enhancing compounds'', and ``amyloid beta'', etc. These recommended queries provide more specific information than \emph{Alzheimer} for deeper and more accurate exploration.
    \vspace{-0.4cm}
\end{itemize}

\vspace{-0.2cm}
\section{Live and Interactive Part}

In the demo session, we will present a live demo and video to explore a AD topic using our system, focusing on the queries of ``\emph{Alzheimer}''. Comprehensive instructions on how to utilize our system will be provided, along with a selection of suggested queries to explore, such as ``\emph{CSF biomarkers}'' and ``\emph{MCI}''.

\section{Conclusion and Future Work}
In this work, we build \sysnameabbr{}, a KG-based retrieval system of biomedical data and articles, aiming to assist scientists in exploring interdisciplinary knowledge in biomedical research. The goal of \sysnameabbr{} is to promote interdisciplinary exploration and improve the understanding of diseases, ultimately benefiting patients.
\sysnameabbr{} is a full-stack system with effective KG-based retrieval, query recommendations, and friendly graphic user interaction.
In the future, we will keep updating this system, including extending the biomedical databases, exploring more effective ways of knowledge graph construction, developing more efficient approaches for query recommendation~\cite{chuang2020tpr}, explainable and fairness demonstrations~\cite{chuang2023efficient,chang2023towards,chuang2022mitigating,jiang2022fmp, yuan2023llm}, and data-centric AI methods~\cite{zha2023data-centric-survey,zha2023data-centric-perspectives}.

\section*{Acknowledgements}
XJ is CPRIT Scholar in Cancer Research (RR180012), and he was supported in part by Christopher Sarofim Family Professorship, UT Stars award, UTHealth startup, the National Institute of Health (NIH) under award number R01AG066749, R01AG066749-03S1, \hash{R01LM013712}, U24LM013755, U01TR002062, U01CA274576, \hash{U54HG012510} and the National Science Foundation (NSF) \#2124789. The authors would like to thank Chen-Chuan Chang and Dr. Chih-Ming Chen for their generous feedback and assistance.

\bibliographystyle{ACM-Reference-Format}
\balance
\bibliography{paper}


\begin{thebibliography}{36}


\ifx \showCODEN    \undefined \def \showCODEN     #1{\unskip}     \fi
\ifx \showDOI      \undefined \def \showDOI       #1{#1}\fi
\ifx \showISBNx    \undefined \def \showISBNx     #1{\unskip}     \fi
\ifx \showISBNxiii \undefined \def \showISBNxiii  #1{\unskip}     \fi
\ifx \showISSN     \undefined \def \showISSN      #1{\unskip}     \fi
\ifx \showLCCN     \undefined \def \showLCCN      #1{\unskip}     \fi
\ifx \shownote     \undefined \def \shownote      #1{#1}          \fi
\ifx \showarticletitle \undefined \def \showarticletitle #1{#1}   \fi
\ifx \showURL      \undefined \def \showURL       {\relax}        \fi
\providecommand\bibfield[2]{#2}
\providecommand\bibinfo[2]{#2}
\providecommand\natexlab[1]{#1}
\providecommand\showeprint[2][]{arXiv:#2}

\bibitem[Archambault et~al\mbox{.}(2011)]%
        {archambault2011difference}
\bibfield{author}{\bibinfo{person}{Daniel Archambault},
  \bibinfo{person}{Helen~C Purchase}, {and} \bibinfo{person}{Bruno Pinaud}.}
  \bibinfo{year}{2011}\natexlab{}.
\newblock \showarticletitle{Difference map readability for dynamic graphs}. In
  \bibinfo{booktitle}{\emph{Graph Drawing: 18th International Symposium, GD
  2010, Konstanz, Germany, September 21-24, 2010. Revised Selected Papers 18}}.
  Springer, \bibinfo{pages}{50--61}.
\newblock


\bibitem[Benson et~al\mbox{.}(2022)]%
        {benson2022don}
\bibfield{author}{\bibinfo{person}{Gloria~S Benson}, \bibinfo{person}{Chris
  Bauer}, \bibinfo{person}{Lucrezia Hausner}, \bibinfo{person}{Samuel
  Couturier}, \bibinfo{person}{Piotr Lewczuk}, \bibinfo{person}{Oliver Peters},
  \bibinfo{person}{Michael H{\"u}ll}, \bibinfo{person}{Holger Jahn},
  \bibinfo{person}{Frank Jessen}, \bibinfo{person}{Johannes Pantel},
  {et~al\mbox{.}}} \bibinfo{year}{2022}\natexlab{}.
\newblock \showarticletitle{Don’t forget about tau: the effects of ApoE4
  genotype on Alzheimer’s disease cerebrospinal fluid biomarkers in subjects
  with mild cognitive impairment—data from the Dementia Competence Network}.
\newblock \bibinfo{journal}{\emph{Journal of Neural Transmission}}
  \bibinfo{volume}{129}, \bibinfo{number}{5-6} (\bibinfo{year}{2022}),
  \bibinfo{pages}{477--486}.
\newblock


\bibitem[Burch and Weiskopf(2014)]%
        {burch2014flip}
\bibfield{author}{\bibinfo{person}{Michael Burch} {and} \bibinfo{person}{Daniel
  Weiskopf}.} \bibinfo{year}{2014}\natexlab{}.
\newblock \showarticletitle{A flip-book of edge-splatted small multiples for
  visualizing dynamic graphs}. In \bibinfo{booktitle}{\emph{Proceedings of the
  7th International Symposium on Visual Information Communication and
  Interaction}}. \bibinfo{pages}{29--38}.
\newblock


\bibitem[Chang et~al\mbox{.}(2020)]%
        {chang2020query}
\bibfield{author}{\bibinfo{person}{Chia-Yuan Chang}, \bibinfo{person}{Ning
  Chen}, \bibinfo{person}{Wei-Ting Chiang}, \bibinfo{person}{Chih-Hen Lee},
  \bibinfo{person}{Yu-Hsuan Tseng}, \bibinfo{person}{Chuan-Ju Wang},
  \bibinfo{person}{Hsien-Hao Chen}, {and} \bibinfo{person}{Ming-Feng Tsai}.}
  \bibinfo{year}{2020}\natexlab{}.
\newblock \showarticletitle{Query Expansion with Semantic-Based Ellipsis
  Reduction for Conversational IR.}. In \bibinfo{booktitle}{\emph{TREC}}.
\newblock


\bibitem[Chang et~al\mbox{.}(2023)]%
        {chang2023towards}
\bibfield{author}{\bibinfo{person}{Chia-Yuan Chang}, \bibinfo{person}{Yu-Neng
  Chuang}, \bibinfo{person}{Kwei-Herng Lai}, \bibinfo{person}{Xiaotian Han},
  \bibinfo{person}{Xia Hu}, {and} \bibinfo{person}{Na Zou}.}
  \bibinfo{year}{2023}\natexlab{}.
\newblock \showarticletitle{Towards Assumption-free Bias Mitigation}.
\newblock \bibinfo{journal}{\emph{arXiv preprint arXiv:2307.04105}}
  (\bibinfo{year}{2023}).
\newblock


\bibitem[Chen et~al\mbox{.}(2016)]%
        {chen2016query}
\bibfield{author}{\bibinfo{person}{Chih-Ming Chen}, \bibinfo{person}{Ming-Feng
  Tsai}, \bibinfo{person}{Yu-Ching Lin}, {and} \bibinfo{person}{Yi-Hsuan
  Yang}.} \bibinfo{year}{2016}\natexlab{}.
\newblock \showarticletitle{Query-based music recommendations via preference
  embedding}. In \bibinfo{booktitle}{\emph{Proceedings of the 10th ACM
  Conference on Recommender Systems}}. \bibinfo{pages}{79--82}.
\newblock


\bibitem[Chu et~al\mbox{.}(2021)]%
        {chu2021cognitive}
\bibfield{author}{\bibinfo{person}{Che-Sheng Chu}, \bibinfo{person}{Cheng-Ta
  Li}, \bibinfo{person}{Andre~R Brunoni}, \bibinfo{person}{Fu-Chi Yang},
  \bibinfo{person}{Ping-Tao Tseng}, \bibinfo{person}{Yu-Kang Tu},
  \bibinfo{person}{Brendon Stubbs}, \bibinfo{person}{Andr{\'e}~F Carvalho},
  \bibinfo{person}{Trevor Thompson}, \bibinfo{person}{Ta-Chuan Yeh},
  {et~al\mbox{.}}} \bibinfo{year}{2021}\natexlab{}.
\newblock \showarticletitle{Cognitive effects and acceptability of non-invasive
  brain stimulation on Alzheimer’s disease and mild cognitive impairment: a
  component network meta-analysis}.
\newblock \bibinfo{journal}{\emph{Journal of Neurology, Neurosurgery \&
  Psychiatry}} \bibinfo{volume}{92}, \bibinfo{number}{2}
  (\bibinfo{year}{2021}), \bibinfo{pages}{195--203}.
\newblock


\bibitem[Chuang et~al\mbox{.}(2020)]%
        {chuang2020tpr}
\bibfield{author}{\bibinfo{person}{Yu-Neng Chuang}, \bibinfo{person}{Chih-Ming
  Chen}, \bibinfo{person}{Chuan-Ju Wang}, \bibinfo{person}{Ming-Feng Tsai},
  \bibinfo{person}{Yuan Fang}, {and} \bibinfo{person}{Ee-Peng Lim}.}
  \bibinfo{year}{2020}\natexlab{}.
\newblock \showarticletitle{TPR: Text-aware preference ranking for recommender
  systems}. In \bibinfo{booktitle}{\emph{Proceedings of the 29th ACM
  International Conference on Information \& Knowledge Management}}.
  \bibinfo{pages}{215--224}.
\newblock


\bibitem[Chuang et~al\mbox{.}(2022)]%
        {chuang2022mitigating}
\bibfield{author}{\bibinfo{person}{Yu-Neng Chuang}, \bibinfo{person}{Kwei-Herng
  Lai}, \bibinfo{person}{Ruixiang Tang}, \bibinfo{person}{Mengnan Du},
  \bibinfo{person}{Chia-Yuan Chang}, \bibinfo{person}{Na Zou}, {and}
  \bibinfo{person}{Xia Hu}.} \bibinfo{year}{2022}\natexlab{}.
\newblock \showarticletitle{Mitigating relational bias on knowledge graphs}.
\newblock \bibinfo{journal}{\emph{arXiv preprint arXiv:2211.14489}}
  (\bibinfo{year}{2022}).
\newblock


\bibitem[Chuang et~al\mbox{.}(2023)]%
        {chuang2023efficient}
\bibfield{author}{\bibinfo{person}{Yu-Neng Chuang}, \bibinfo{person}{Guanchu
  Wang}, \bibinfo{person}{Fan Yang}, \bibinfo{person}{Zirui Liu},
  \bibinfo{person}{Xuanting Cai}, \bibinfo{person}{Mengnan Du}, {and}
  \bibinfo{person}{Xia Hu}.} \bibinfo{year}{2023}\natexlab{}.
\newblock \showarticletitle{Efficient xai techniques: A taxonomic survey}.
\newblock \bibinfo{journal}{\emph{arXiv preprint arXiv:2302.03225}}
  (\bibinfo{year}{2023}).
\newblock


\bibitem[Dong et~al\mbox{.}(2023a)]%
        {dong2023hierarchy}
\bibfield{author}{\bibinfo{person}{Junnan Dong}, \bibinfo{person}{Qinggang
  Zhang}, \bibinfo{person}{Xiao Huang}, \bibinfo{person}{Keyu Duan},
  \bibinfo{person}{Qiaoyu Tan}, {and} \bibinfo{person}{Zhimeng Jiang}.}
  \bibinfo{year}{2023}\natexlab{a}.
\newblock \showarticletitle{Hierarchy-Aware Multi-Hop Question Answering over
  Knowledge Graphs}. In \bibinfo{booktitle}{\emph{Proceedings of the ACM Web
  Conference 2023}}. \bibinfo{pages}{2519--2527}.
\newblock


\bibitem[Dong et~al\mbox{.}(2023b)]%
        {dong2023active}
\bibfield{author}{\bibinfo{person}{Junnan Dong}, \bibinfo{person}{Qinggang
  Zhang}, \bibinfo{person}{Xiao Huang}, \bibinfo{person}{Qiaoyu Tan},
  \bibinfo{person}{Daochen Zha}, {and} \bibinfo{person}{Zhao Zihao}.}
  \bibinfo{year}{2023}\natexlab{b}.
\newblock \showarticletitle{Active ensemble learning for knowledge graph error
  detection}. In \bibinfo{booktitle}{\emph{Proceedings of the Sixteenth ACM
  International Conference on Web Search and Data Mining}}.
  \bibinfo{pages}{877--885}.
\newblock


\bibitem[Hogan et~al\mbox{.}(2021)]%
        {hogan2021knowledge}
\bibfield{author}{\bibinfo{person}{Aidan Hogan}, \bibinfo{person}{Eva
  Blomqvist}, \bibinfo{person}{Michael Cochez}, \bibinfo{person}{Claudia
  d’Amato}, \bibinfo{person}{Gerard~de Melo}, \bibinfo{person}{Claudio
  Gutierrez}, \bibinfo{person}{Sabrina Kirrane}, \bibinfo{person}{Jos{\'e}
  Emilio~Labra Gayo}, \bibinfo{person}{Roberto Navigli},
  \bibinfo{person}{Sebastian Neumaier}, {et~al\mbox{.}}}
  \bibinfo{year}{2021}\natexlab{}.
\newblock \showarticletitle{Knowledge graphs}.
\newblock \bibinfo{journal}{\emph{ACM Computing Surveys (CSUR)}}
  \bibinfo{volume}{54}, \bibinfo{number}{4} (\bibinfo{year}{2021}),
  \bibinfo{pages}{1--37}.
\newblock


\bibitem[Jiang et~al\mbox{.}(2022)]%
        {jiang2022fmp}
\bibfield{author}{\bibinfo{person}{Zhimeng Jiang}, \bibinfo{person}{Xiaotian
  Han}, \bibinfo{person}{Chao Fan}, \bibinfo{person}{Zirui Liu},
  \bibinfo{person}{Na Zou}, \bibinfo{person}{Ali Mostafavi}, {and}
  \bibinfo{person}{Xia Hu}.} \bibinfo{year}{2022}\natexlab{}.
\newblock \showarticletitle{Fmp: Toward fair graph message passing against
  topology bias}.
\newblock \bibinfo{journal}{\emph{arXiv preprint arXiv:2202.04187}}
  (\bibinfo{year}{2022}).
\newblock


\bibitem[Lee et~al\mbox{.}(2020)]%
        {lee2020biobert}
\bibfield{author}{\bibinfo{person}{Jinhyuk Lee}, \bibinfo{person}{Wonjin Yoon},
  \bibinfo{person}{Sungdong Kim}, \bibinfo{person}{Donghyeon Kim},
  \bibinfo{person}{Sunkyu Kim}, \bibinfo{person}{Chan~Ho So}, {and}
  \bibinfo{person}{Jaewoo Kang}.} \bibinfo{year}{2020}\natexlab{}.
\newblock \showarticletitle{BioBERT: a pre-trained biomedical language
  representation model for biomedical text mining}.
\newblock \bibinfo{journal}{\emph{Bioinformatics}} \bibinfo{volume}{36},
  \bibinfo{number}{4} (\bibinfo{year}{2020}), \bibinfo{pages}{1234--1240}.
\newblock


\bibitem[Liu et~al\mbox{.}(2022)]%
        {liu2022ipr}
\bibfield{author}{\bibinfo{person}{Shih-Yang Liu}, \bibinfo{person}{Hsien~Hao
  Chen}, \bibinfo{person}{Chih-Ming Chen}, \bibinfo{person}{Ming-Feng Tsai},
  {and} \bibinfo{person}{Chuan-Ju Wang}.} \bibinfo{year}{2022}\natexlab{}.
\newblock \showarticletitle{IPR: Interaction-level Preference Ranking for
  Explicit Feedback}. In \bibinfo{booktitle}{\emph{Proceedings of the 45th
  International ACM SIGIR Conference on Research and Development in Information
  Retrieval}}. \bibinfo{pages}{1912--1916}.
\newblock


\bibitem[Marcelli et~al\mbox{.}(2018)]%
        {marcelli2018involvement}
\bibfield{author}{\bibinfo{person}{Serena Marcelli}, \bibinfo{person}{Massimo
  Corbo}, \bibinfo{person}{Filomena Iannuzzi}, \bibinfo{person}{Lucia Negri},
  \bibinfo{person}{Fabio Blandini}, \bibinfo{person}{Robert Nistico}, {and}
  \bibinfo{person}{Marco Feligioni}.} \bibinfo{year}{2018}\natexlab{}.
\newblock \showarticletitle{The involvement of post-translational modifications
  in Alzheimer's disease}.
\newblock \bibinfo{journal}{\emph{Current Alzheimer Research}}
  \bibinfo{volume}{15}, \bibinfo{number}{4} (\bibinfo{year}{2018}),
  \bibinfo{pages}{313--335}.
\newblock


\bibitem[Meta(2013)]%
        {react}
\bibfield{author}{\bibinfo{person}{Meta}.} \bibinfo{year}{2013}\natexlab{}.
\newblock \bibinfo{title}{React}.
\newblock
\newblock
\urldef\tempurl%
\url{http://reactjs.org}
\showURL{%
Retrieved Feb 15th, 2023 from \tempurl}


\bibitem[Neo4j(2007)]%
        {neo4j}
\bibfield{author}{\bibinfo{person}{Neo4j}.} \bibinfo{year}{2007}\natexlab{}.
\newblock \bibinfo{title}{Neo4j}.
\newblock
\newblock
\urldef\tempurl%
\url{https://neo4j.com}
\showURL{%
Retrieved Feb 15th, 2023 from \tempurl}


\bibitem[Nicholson and Greene(2020)]%
        {nicholson2020constructing}
\bibfield{author}{\bibinfo{person}{David~N Nicholson} {and}
  \bibinfo{person}{Casey~S Greene}.} \bibinfo{year}{2020}\natexlab{}.
\newblock \showarticletitle{Constructing knowledge graphs and their biomedical
  applications}.
\newblock \bibinfo{journal}{\emph{Computational and structural biotechnology
  journal}}  \bibinfo{volume}{18} (\bibinfo{year}{2020}),
  \bibinfo{pages}{1414--1428}.
\newblock


\bibitem[Ronacher(2010)]%
        {flask}
\bibfield{author}{\bibinfo{person}{Armin Ronacher}.}
  \bibinfo{year}{2010}\natexlab{}.
\newblock \bibinfo{title}{Flask}.
\newblock
\newblock
\urldef\tempurl%
\url{https://flask.palletsprojects.com/en/2.2.x/}
\showURL{%
Retrieved Feb 15th, 2023 from \tempurl}


\bibitem[Srinivasa-Desikan(2018)]%
        {srinivasa2018natural}
\bibfield{author}{\bibinfo{person}{Bhargav Srinivasa-Desikan}.}
  \bibinfo{year}{2018}\natexlab{}.
\newblock \bibinfo{booktitle}{\emph{Natural Language Processing and
  Computational Linguistics: A practical guide to text analysis with Python,
  Gensim, spaCy, and Keras}}.
\newblock \bibinfo{publisher}{Packt Publishing Ltd}.
\newblock


\bibitem[Stewart et~al\mbox{.}(2019)]%
        {stewart2019icdm}
\bibfield{author}{\bibinfo{person}{Michael Stewart},
  \bibinfo{person}{Majigsuren Enkhsaikhan}, {and} \bibinfo{person}{Wei Liu}.}
  \bibinfo{year}{2019}\natexlab{}.
\newblock \showarticletitle{Icdm 2019 knowledge graph contest: Team uwa}. In
  \bibinfo{booktitle}{\emph{2019 IEEE international conference on data mining
  (ICDM)}}. IEEE, \bibinfo{pages}{1546--1551}.
\newblock


\bibitem[Tan et~al\mbox{.}(2019)]%
        {tan2019deep}
\bibfield{author}{\bibinfo{person}{Qiaoyu Tan}, \bibinfo{person}{Ninghao Liu},
  {and} \bibinfo{person}{Xia Hu}.} \bibinfo{year}{2019}\natexlab{}.
\newblock \showarticletitle{Deep representation learning for social network
  analysis}.
\newblock \bibinfo{journal}{\emph{Frontiers in big Data}}  \bibinfo{volume}{2}
  (\bibinfo{year}{2019}), \bibinfo{pages}{2}.
\newblock


\bibitem[Tan et~al\mbox{.}(2020)]%
        {tan2020learning}
\bibfield{author}{\bibinfo{person}{Qiaoyu Tan}, \bibinfo{person}{Ninghao Liu},
  \bibinfo{person}{Xing Zhao}, \bibinfo{person}{Hongxia Yang},
  \bibinfo{person}{Jingren Zhou}, {and} \bibinfo{person}{Xia Hu}.}
  \bibinfo{year}{2020}\natexlab{}.
\newblock \showarticletitle{Learning to hash with graph neural networks for
  recommender systems}. In \bibinfo{booktitle}{\emph{Proceedings of The Web
  Conference 2020}}. \bibinfo{pages}{1988--1998}.
\newblock


\bibitem[Tan et~al\mbox{.}(2023a)]%
        {tan2023collaborative}
\bibfield{author}{\bibinfo{person}{Qiaoyu Tan}, \bibinfo{person}{Xin Zhang},
  \bibinfo{person}{Xiao Huang}, \bibinfo{person}{Hao Chen},
  \bibinfo{person}{Jundong Li}, {and} \bibinfo{person}{Xia Hu}.}
  \bibinfo{year}{2023}\natexlab{a}.
\newblock \showarticletitle{Collaborative Graph Neural Networks for Attributed
  Network Embedding}.
\newblock \bibinfo{journal}{\emph{IEEE Transactions on Knowledge and Data
  Engineering}} (\bibinfo{year}{2023}).
\newblock


\bibitem[Tan et~al\mbox{.}(2023b)]%
        {tan2023bring}
\bibfield{author}{\bibinfo{person}{Qiaoyu Tan}, \bibinfo{person}{Xin Zhang},
  \bibinfo{person}{Ninghao Liu}, \bibinfo{person}{Daochen Zha},
  \bibinfo{person}{Li Li}, \bibinfo{person}{Rui Chen},
  \bibinfo{person}{Soo-Hyun Choi}, {and} \bibinfo{person}{Xia Hu}.}
  \bibinfo{year}{2023}\natexlab{b}.
\newblock \showarticletitle{Bring your own view: Graph neural networks for link
  prediction with personalized subgraph selection}. In
  \bibinfo{booktitle}{\emph{Proceedings of the Sixteenth ACM International
  Conference on Web Search and Data Mining}}. \bibinfo{pages}{625--633}.
\newblock


\bibitem[Wang et~al\mbox{.}(2019)]%
        {wang2019kgat}
\bibfield{author}{\bibinfo{person}{Xiang Wang}, \bibinfo{person}{Xiangnan He},
  \bibinfo{person}{Yixin Cao}, \bibinfo{person}{Meng Liu}, {and}
  \bibinfo{person}{Tat-Seng Chua}.} \bibinfo{year}{2019}\natexlab{}.
\newblock \showarticletitle{Kgat: Knowledge graph attention network for
  recommendation}. In \bibinfo{booktitle}{\emph{Proceedings of the 25th ACM
  SIGKDD international conference on knowledge discovery \& data mining}}.
  \bibinfo{pages}{950--958}.
\newblock


\bibitem[Wu et~al\mbox{.}(2023)]%
        {wu2023medical}
\bibfield{author}{\bibinfo{person}{Xuehong Wu}, \bibinfo{person}{Junwen Duan},
  \bibinfo{person}{Yi Pan}, {and} \bibinfo{person}{Min Li}.}
  \bibinfo{year}{2023}\natexlab{}.
\newblock \showarticletitle{Medical knowledge graph: Data sources,
  construction, reasoning, and applications}.
\newblock \bibinfo{journal}{\emph{Big Data Mining and Analytics}}
  \bibinfo{volume}{6}, \bibinfo{number}{2} (\bibinfo{year}{2023}),
  \bibinfo{pages}{201--217}.
\newblock


\bibitem[Xu et~al\mbox{.}(2021)]%
        {xu2021inhibition}
\bibfield{author}{\bibinfo{person}{Lu Xu}, \bibinfo{person}{Cai-Long Pan},
  \bibinfo{person}{Xiang-Hui Wu}, \bibinfo{person}{Jing-Jing Song},
  \bibinfo{person}{Ping Meng}, \bibinfo{person}{Lei Li}, \bibinfo{person}{Li
  Wang}, \bibinfo{person}{Zhiren Zhang}, {and} \bibinfo{person}{Zhi-Yuan
  Zhang}.} \bibinfo{year}{2021}\natexlab{}.
\newblock \showarticletitle{Inhibition of Smad3 in macrophages promotes
  A$\beta$ efflux from the brain and thereby ameliorates Alzheimer's
  pathology}.
\newblock \bibinfo{journal}{\emph{Brain, Behavior, and Immunity}}
  \bibinfo{volume}{95} (\bibinfo{year}{2021}), \bibinfo{pages}{154--167}.
\newblock


\bibitem[Ye et~al\mbox{.}(2022)]%
        {ye2022generative}
\bibfield{author}{\bibinfo{person}{Hongbin Ye}, \bibinfo{person}{Ningyu Zhang},
  \bibinfo{person}{Hui Chen}, {and} \bibinfo{person}{Huajun Chen}.}
  \bibinfo{year}{2022}\natexlab{}.
\newblock \showarticletitle{Generative knowledge graph construction: A review}.
\newblock \bibinfo{journal}{\emph{arXiv preprint arXiv:2210.12714}}
  (\bibinfo{year}{2022}).
\newblock


\bibitem[Yuan et~al\mbox{.}(2023)]%
        {yuan2023llm}
\bibfield{author}{\bibinfo{person}{Jiayi Yuan}, \bibinfo{person}{Ruixiang
  Tang}, \bibinfo{person}{Xiaoqian Jiang}, {and} \bibinfo{person}{Xia Hu}.}
  \bibinfo{year}{2023}\natexlab{}.
\newblock \showarticletitle{LLM for Patient-Trial Matching: Privacy-Aware Data
  Augmentation Towards Better Performance and Generalizability}.
\newblock \bibinfo{journal}{\emph{arXiv preprint arXiv:2303.16756}}
  (\bibinfo{year}{2023}).
\newblock


\bibitem[Zha et~al\mbox{.}(2023a)]%
        {zha2023data-centric-perspectives}
\bibfield{author}{\bibinfo{person}{Daochen Zha}, \bibinfo{person}{Zaid~Pervaiz
  Bhat}, \bibinfo{person}{Kwei-Herng Lai}, \bibinfo{person}{Fan Yang}, {and}
  \bibinfo{person}{Xia Hu}.} \bibinfo{year}{2023}\natexlab{a}.
\newblock \showarticletitle{Data-centric AI: Perspectives and Challenges}. In
  \bibinfo{booktitle}{\emph{Proceedings of the 2023 SIAM International
  Conference on Data Mining (SDM)}}.
\newblock


\bibitem[Zha et~al\mbox{.}(2023b)]%
        {zha2023data-centric-survey}
\bibfield{author}{\bibinfo{person}{Daochen Zha}, \bibinfo{person}{Zaid~Pervaiz
  Bhat}, \bibinfo{person}{Kwei-Herng Lai}, \bibinfo{person}{Fan Yang},
  \bibinfo{person}{Zhimeng Jiang}, \bibinfo{person}{Shaochen Zhong}, {and}
  \bibinfo{person}{Xia Hu}.} \bibinfo{year}{2023}\natexlab{b}.
\newblock \showarticletitle{Data-centric Artificial Intelligence: A Survey}.
\newblock \bibinfo{journal}{\emph{arXiv preprint arXiv:2303.10158}}
  (\bibinfo{year}{2023}).
\newblock


\bibitem[Zhou et~al\mbox{.}(2023)]%
        {zhou2023adaptive}
\bibfield{author}{\bibinfo{person}{Huachi Zhou}, \bibinfo{person}{Hao Chen},
  \bibinfo{person}{Junnan Dong}, \bibinfo{person}{Daochen Zha},
  \bibinfo{person}{Chuang Zhou}, {and} \bibinfo{person}{Xiao Huang}.}
  \bibinfo{year}{2023}\natexlab{}.
\newblock \showarticletitle{Adaptive Popularity Debiasing Aggregator for Graph
  Collaborative Filtering}. In \bibinfo{booktitle}{\emph{Proceedings of the
  46th International ACM SIGIR Conference on Research and Development in
  Information Retrieval}}.
\newblock


\bibitem[Zhu et~al\mbox{.}(2022)]%
        {zhu2022multi}
\bibfield{author}{\bibinfo{person}{Xiangru Zhu}, \bibinfo{person}{Zhixu Li},
  \bibinfo{person}{Xiaodan Wang}, \bibinfo{person}{Xueyao Jiang},
  \bibinfo{person}{Penglei Sun}, \bibinfo{person}{Xuwu Wang},
  \bibinfo{person}{Yanghua Xiao}, {and} \bibinfo{person}{Nicholas~Jing Yuan}.}
  \bibinfo{year}{2022}\natexlab{}.
\newblock \showarticletitle{Multi-modal knowledge graph construction and
  application: A survey}.
\newblock \bibinfo{journal}{\emph{IEEE Transactions on Knowledge and Data
  Engineering}} (\bibinfo{year}{2022}).
\newblock


\end{thebibliography}

\end{document}